\documentclass[review]{elsarticle}

\usepackage{graphicx}
\usepackage{epstopdf}
\usepackage{subcaption}
\usepackage{color}
\usepackage{rotating}

\journal{Journal of \LaTeX\ Templates}

\begin{document}

\begin{frontmatter}



\title{Plasma streams in the Hermean dayside magnetosphere: solar wind injection through the reconnection region}


\author[label1,label2,label3]{J. Varela}
\ead{deviriam@gmail.com (telf: 0033782822476)}

\author[label3]{F. Pantellini}
\author[label3]{M. Moncuquet}

\address[label1]{LIMSI, CNRS, Orsay, France}
\address[label2]{AIM DSM/IRFU/SAp, CEA Saclay, France}
\address[label3]{LESIA, Observatoire de Paris, CNRS, UPMC, Universite Paris-Diderot, 5
place Jules Janssen, 92195 Meudon, France}

\begin{abstract}

The aim of this research is to simulate the interaction of the solar wind with the magnetic field of Mercury and to study the particle fluxes between the magnetosheath and the planet surface. We simulate the magnetosphere structure using the open source MHD code PLUTO in spherical geometry with a multipolar expansion of the Hermean magnetic field (Anderson, B. J. et al, 2012). We perform two simulations with realistic solar wind parameters to study the properties of a plasma stream originated in the reconnection region between the interplanetary and the Hermean magnetic field. The plasma precipitates along the open magnetic field lines to the planet surface showing a fast expansion, rarefaction and cooling. The plasma stream is correlated with a flattening of the magnetic field observed by MESSENGER due to the adjacency of the reconnection region where the solar wind is injected to the inner magnetosphere.

\end{abstract}


\end{frontmatter}


\section{Introduction}
\label{Introduction}

Mercury is the closest planet to the Sun ($0.47$ AU in the Aphelion and $0.31$ AU in the Perihelion), it has the smallest mean radius ($R_{M} = 2439.7$ km), it is the most dense ($5.427$ cm$^{-3}$) and with the most eccentric orbit ($0.206$) among the telluric planets. The internal structure of Mercury is not well known and there are several theories that try to explain the origin of its global magnetic field, very unusual characteristic due to its small size and slow rotation (59 days) \cite{2003EandPSL.208....1S}. The most suitable explanation of the magnetic field origin is the presence of a molten core \cite{2007Sci...316..710M}. There are several models that try to explain the observations, like the thin-shell dynamo \cite{2006GeoRL..3310202T}, the deep dynamo \cite{2005EandPSL.236..542H}, deep dynamos enclosed by a stably stratified electrically conductive layer \cite{2010EandPSL.289..619M}, induction feedback on the convecting portion of the core from magnetopause currents \cite{Heyner1690} or the precipitation of solid iron in radial zones within a liquid outer core \cite{JGRE:JGRE2765}.

The spacecraft Mariner 10 visited Mercury in three flybys (29 March 1974, 21 September 1974 and 16 March 1975) and revealed an intrinsic magnetic field measuring 400 nT during the closest approach at 327 km, as well as a variable magnetosphere (MS) and a bow shock (BS) \cite{1974Sci...185..131N,1979JGR....84.2076S}. The analysis of the data points out that the Hermean magnetic field is a dipole \cite{1975JGR....80.2708N}. The second spacecraft to visit Mercury is MESSENGER with three flybys (14 January 2008, 06 October 2008 and 29 September 2009) and the orbital insertion on 18 March 2011 \cite{2001PandSS...49.1445S,2008Sci...321...85S,Anderson2010}. The data analysis revealed a dipole shifted northward by $479 \pm 6$ km, a dipolar moment of $195$ nT $\cdot R^{3}_{M}$ and a tilt of the magnetic axis relative to the planetary spin axis smaller than $0.8^{0}$ \cite{2011Sci...333.1859A}. The new data from more than a thousand of orbits in the North hemisphere showed that the Hermean magnetic field can be modeled by an axisymmetric multipolar expansion \cite{2012JGRA..11710228R,2012JGRE..117.0L12A}.

MESSENGER magnetometer data shows the large variability of the Herman magnetosphere as a function of the interplanetary magnetic field (IMF). The range of values observed for the IMF module goes from less than $10$ nT to more than $60$ nT, with orientations sometimes strongly departing from the Parker spiral. Other parameters of the solar wind (SW) are estimated using numerical models. The density can oscillate between $30$ to $160$ cm$^{-3}$, from $45,000$ to $160,000$ K for the temperature, between $250$ to $600$ km/s for the velocity and $\beta$ values from $0.08$ to $1$ \cite{2009JGRA..11410101B,2011PandSS...59.2066B,2013JGRA..118...45B,2008Sci...321...82A,JGRE:JGRE3136}.

In the present paper we use a single fluid MHD model to study the global structure of the Hermean magnetosphere for realistic conditions of the solar wind parameters. We use the code PLUTO \cite{2007ApJS..170..228M} in 3D spherical coordinates without resistivity and viscosity. We use realistic parameters for the SW obtained by the numerical models ENLIL + GONG WSA + Cone SWRC \cite{ODSTRCIL2003497,SWE:SWE449} and the IMF data from the MESSENGER magnetometer.

Present study is devoted to complement previous observational studies that revealed the presence of a thick plasma depletion layer in the magnetosheath between the subsolar magnetopause and the bow shock \cite{2013JGRA..118.7181G,2013AGUFMSM24A..03D}, as well as other theoretical studies dedicated to the simulation of the global structures of the Hermean magnetosphere using MHD \cite{2008Icar..195....1K} and Hybrid \cite{2007AGUFMSM53C1412T,2010Icar..209...11T} numerical models.

In a previous communication the authors analyzed the effect of the IMF orientation in the Hermean magnetosphere and the plasma flows toward the planet surface \cite{2015PandSS..119..264V}. A thorough study of the plasma flows properties was avoided, we only indicated that the local maximum of the inflow on the planet surface is displaced when the IMF orientation changes as well as the location of the reconnection region. The aim of the present research is to improve the analysis of the plasma flows and define the concept of plasma stream. We study the plasma stream characteristics from its origin at the magnetosheath to the final precipitation on the planet surface, adding the expected integrated value and the distribution of the mass deposition as well as the particle sputtering on the planet surface. We show two simulations with different SW conditions, leading to distinctive magnetosphere configurations and plasma stream properties, modifying the mass deposition and particle sputtering distributions.

This paper is structured as follows. In Section 2, a description of the simulation model and the initial parameters used are provided. In Section 3, the model results for the MESSENGER orbits. In Section 4, comparison of the simulation results with MESSENGER data. In Section 5, conclusion and discussion.   

\section{Numerical model}
\label{Model}

We use the MHD version of the code PLUTO in spherical coordinates to simulate a single fluid polytropic plasma in the non resistive and inviscid limit. The code is freely available online \cite{2007ApJS..170..228M}.

The simulation domain is confined within two spherical shells centered in the planet, representing the inner and outer boundaries of the system. Between the inner shell and the planet surface (at radius unity in the domain) there is a "soft coupling region" where special conditions apply (defined in the next section).The shells are at $0.6 R_{M}$ and $12 R_{M}$ ($R_{M}$ is the Mercury radius).

The conservative form of the equations are integrated using a Harten, Lax, Van Leer approximate Riemann solver (hll) associated with a diffusive limiter (minmod). The divergence of the magnetic field is ensured by a mixed hyperbolic/parabolic divergence cleaning technique (DIV CLEANING) \cite{2002JCoPh.175..645D}.

The grid points are $196$ radial points, $48$ in the polar angle $\theta$ and $96$ in the azimuthal angle $\phi$ (the grid poles correspond to the magnetic poles).

The planetary magnetic field is axisymmetric with the magnetic potential $\Psi$ expanded in dipolar, quadrupolar, octupolar and 16-polar terms \cite{2012JGRE..117.0L12A}:

$$ \Psi (r,\theta) = R_{M}\sum^{4}_{l=1} (\frac{R_{M}}{r})^{l+1} g_{l0} P_{l}(cos\theta) $$

The current free magnetic field is $B_{M} = -\nabla \Psi $. $r$ is the distance to the planet center and $\theta$ the polar angle. The Legendre polynomials in the magnetic potential $\Psi$ are:

$$ P_{1}(x) = x $$
$$ P_{2}(x) = \frac{1}{2} (3x^2 - 1) $$
$$ P_{3}(x) = \frac{1}{2} (5x^3 - 3x) $$
$$ P_{4}(x) = \frac{1}{2} (35x^4 - 30x^2 + 3) $$
the numerical coefficients $g_{l0}$ taken from Anderson et al. 2012 are summarized in the Table 1 \cite{2012JGRE..117.0L12A}.

\begin{table}[h]
\centering
\begin{tabular}{c | c c c c}
coeff & $g_{01}$(nT) & $g_{02}/g_{01}$ & $g_{03}/g_{01}$ & $g_{04}/g_{01}$  \\ \hline
 & $-182$ & $0.4096$ & $0.1265$ & $0.0301$ \\
\end{tabular}
\caption{Multipolar coefficients $g_{l0}$ for Mercury's internal field.}
\end{table}

The simulation frame is such that the z-axis is given by the planetary magnetic axis pointing to the magnetic North pole and the Sun is located in the XZ plane with $x_{sun} > 0$. The y-axis completes the right-handed system.

\subsection{Boundary conditions and initial conditions}

The outer boundary is divided in two regions, the upstream part where the solar wind parameters are fixed and the downstream part where we consider the null derivative condition $\frac{\partial}{\partial r} = 0$ for all fields. At the inner boundary the value of the intrinsic magnetic field of Mercury is specified. In the soft coupling region the velocity is smoothly reduced to zero when approaching the inner boundary. The magnetic field and the velocity are parallel, and the density is adjusted to keep the Alfven velocity constant $v_{A} = B / \sqrt{\mu_{0}\rho} = 25$ km/s with $\rho = nm_{p}$ the mass density, $n$ the particle number, $m_{p}$ the proton mass and $\mu_{0}$ the vacuum magnetic permeability. In the initial conditions we define a paraboloid in the night side with the vertex at the center of the planet where the velocity is null and the density is two order smaller than in the solar wind. The IMF is cut off at $2 R_{M}$.

The solar wind parameters in the simulations are summarized in Table 2. We assume a fully ionized proton electron plasma, the sound speed is defined as $v_{s} = \sqrt {\gamma p/\rho} $ (with $p$ the total electron + proton pressure), the sonic Mach number as $M_{s} = v/v_{s}$ with $v$ the velocity and $M_{A} = v/v_{A}$ the Alfvenic Mach number. $\vec{v}_{u}$ is the unitary vector of the velocity.

\begin{table}[h]
\centering
\begin{tabular}{c | c c c c c }
Name & Date & B field (nT) & n (cm$^{-3}$) & $T$ (K) & $\beta$ \\ \hline
 Orbit I & 2012/06/26 & $(16,-6,10)$ & $15$ & $160000$ & $0.23$ \\
 Orbit II & 2011/11/02 & $(18,-2,5)$ & $20$ & $95000$ & $0.19$ \\
\end{tabular}
\caption{Simulation parameters I}
\end{table}

\begin{table}[h]
\centering
\begin{tabular}{c | c c c c c c c c}
Name & $v$ (km/s) & $\vec{v}_{u}$ & $M_{s}$  & $M_{A}$\\ \hline
 Orbit I & $500$ & $(-0.997,0.070,0)$ &$9.1$ & 4.46\\
 Orbit II & $360$ & $(-0.994,0.110,0)$ & $7.1$ & 3.91\\
\end{tabular}
\caption{Simulations parameters II}
\end{table}

The IMF orientation and the solar wind $\beta$ are similar in the orbits I and II simulations, but the sonic Mach number is larger in the orbit I leading to a stronger compression of the bow shock that will affect the properties of the plasma stream. We analyze and compare both cases in the next section.

\section{Simulation results}
\label{Results}

Figure 1 shows the module of the magnetic field over the magnetic field lines (colored lines) and in a frontal plane at $X = 0.3R_{M}$ for the simulation of the orbits I and II (black line). There are closed magnetic field lines on the day side and open magnetic field lines at high latitudes. The reconnection regions are observed nearby the poles (dark blue color in the frontal plane), wider and displaced to the West (East) in the North (South) Hemisphere in the orbit I simulation because the $Y$ component of the magnetic field is larger than in the orbit II case. The plasma stream lines (green lines) show how the solar wind is injected in the inner magnetosphere across the reconnection region, the precipitation of the plasma stream on the planet surface following the open magnetic field lines near the poles as well as the correlation of the plasma stream with inflow regions on the planet surface (blue color om the planet surface). 

\begin{figure}[h]
\centering
\includegraphics[width=1.0\textwidth]{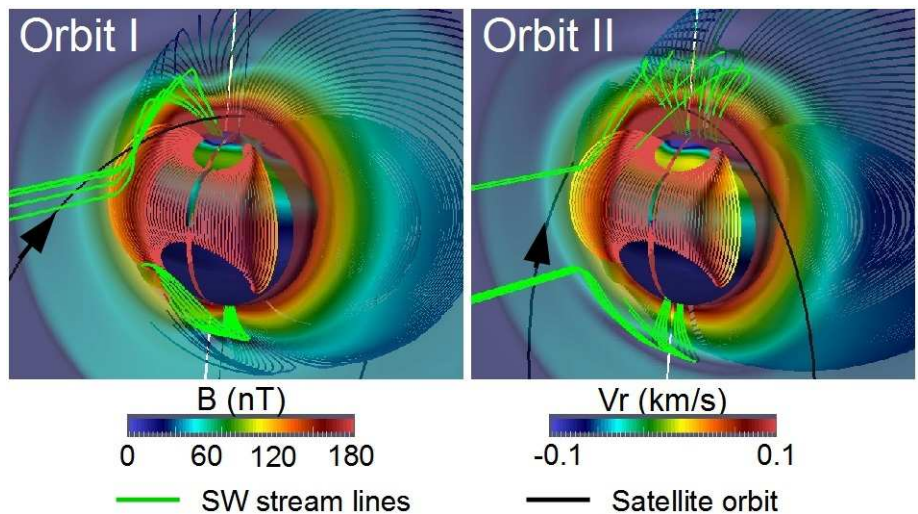}
\caption{Hermean magnetic field lines colored with the module of the magnetic field. Magnetic field module in the frontal plane $X = 0.3R_{M}$. Inflow/outflow regions in the planet surface (blue/red). Satellite trajectory (black). Solar wind stream lines (green). All the stream lines are connected to the solar wind (the lines that crosses the plane $Z = 0$ are artificially cut).}
\end{figure}

Figure 2 shows a polar cut of the density distribution for the orbit I and II simulations. There are three distinct regions, the upstream region of the solar wind, the BS with a sharp increase of the density and the interior of the magnetosphere where the density drops several orders of magnitude. There is a plasma stream that links the back of the BS, the magnetosheath, with the planet surface at the North and South hemispheres, in the interface of closed/open magnetic field lines on the dayside at the Hermean cusp. The plasma stream is wider in the North Hemisphere due to the Northward displacement of the Hermean magnetic field. In the South Hemisphere the back of the BS reaches the planet surface so the plasma stream is more difficult to observe. The white line shows the iso-line of the magnetic field of $50$ nT, indicating that the origin of the plasma stream in the magnetosheath is correlated with a local drop of the magnetic field due to the proximity of the reconnection region.

The pink dot indicates the satellite closest approach of each orbit. The black lines along the plasma streams at both Hemispheres indicate the region plotted in the Figure 3.

\begin{figure}[h]
\centering
\includegraphics[width=0.9\textwidth]{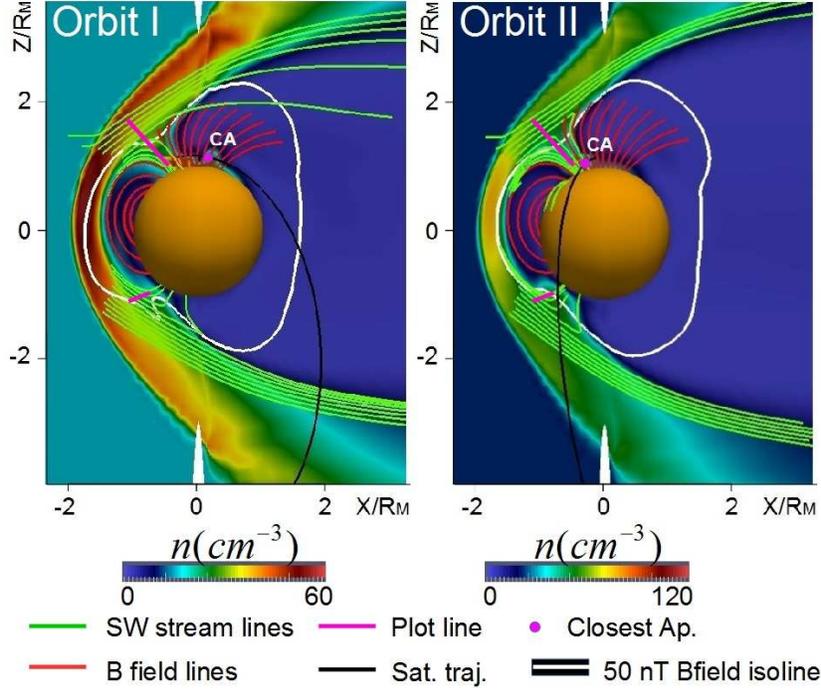}
\caption{Polar cuts of the density distribution in the orbit I and orbit II simulations. Magnetic field lines (red lines) and solar wind stream functions (green lines). The pink dot shows the closest approach (CA) of the satellite. The pink lines show the region plotted in the Figure 3. The black line is the satellite trajectory. The white line indicates the magnetic field iso-line of 50 nT.}
\end{figure}

Figure 3 shows the profiles of the pressure, density, temperature, velocity (module and components) and magnetic field module along the plasma stream structure (pink lines, Figure 2). The plots show the origin of the plasma stream in the magnetosheath (left on the graphs) until its precipitation on the planet surface (right on the graphs). We observe a similar pattern, particularly  in the North pole where the plasma stream structure is more robust: the plasma stream originates in the magnetosheath nearby the reconnection region, identified in the graphs as a flattening of the magnetic field module for the orbit I simulation (from $L = 0.3 R_{M}$ to $0.5 R_{M}$) and as a flattening in between two local drops for the orbit II simulation (from $L = 0.2 R_{M}$ to $0.4 R_{M}$) because the satellite trajectory is closer to the reconnection X point of the reconnection (see figure 1). In this region the plasma is locally hotter, more dense and it is slow down. From the reconnection region a plasma stream precipitates along the open magnetic field lines an the plasma suffer an expansion (increase of the velocity, particularly the $Z$ component), cooling and rarefaction. In both simulations, nearby the North pole, there is a second local maximum of the density indicating a region of cold and dense plasma that it is decelerated and accumulated before precipitate on the planet surface.

\begin{figure}[h]
\centering
\includegraphics[width=0.8\textwidth]{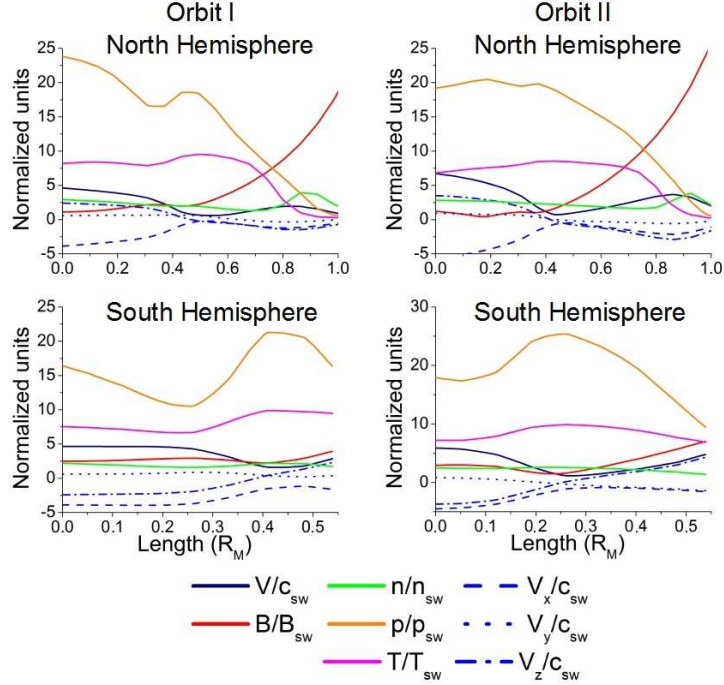}
\caption{Pressure, density, temperature, velocity (module and components) and magnetic field module along the plasma stream (black lines, Figure 2) for the orbit I and II simulations at the North and South Hemispheres. Pressure, density, temperature and magnetic field are normalized to the SW value and the velocity to the SW sound speed.}
\end{figure}

Figure 4 shows the regions of inflow/outflow (blue/red) on the planet surface. In both simulations there are regions of inflow at the North Hemisphere near the poles, and for latitudes larger than $25^{o}$ at the South Hemisphere. The inflow regions at the South Hemisphere are wider than at the North Hemisphere. Comparing simulations, the orbit I simulation there is a stronger inflow at the North Hemisphere and a larger East-West asymmetry than in the orbit II case. The region with open magnetic field lines is wilder in the South Hemisphere (light blue dots). The East-West asymmetry of the Hermean magnetic field is observed too in the open magnetic field lines distribution of the orbit I simulation. The regions with open magnetic field lines are bigger in the orbit II simulation but no East-West asymmetry is observed (the minimum latitude with open magnetic field lines is similar in both cases).

\begin{figure}[h]
\centering
\includegraphics[width=1.0\textwidth]{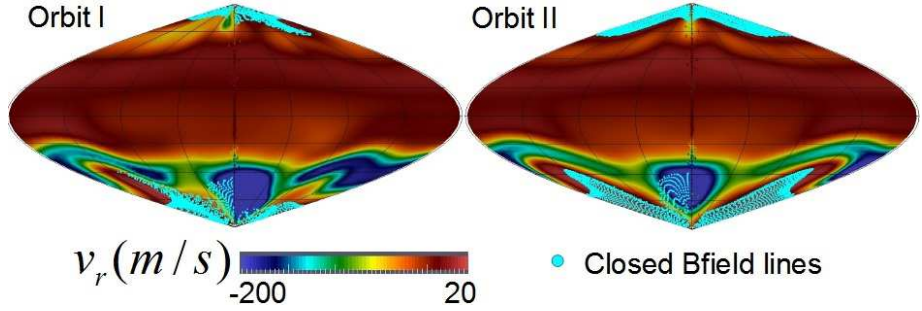}
\caption{Sinusoidal (Sanson-Flamsteed) projection of the inflow-outflow regions (blue-red) in the simulation of the orbits I and II. Open magnetic field lines regions (light blue dots) in the planet surface.}
\end{figure}

Figure 5 shows the mass deposition on the planet surface. The mass deposition in the orbit I simulation is more asymmetric than in the orbit II case due to the West-East displacement of the reconnection regions. The mass deposition on the North Hemisphere is located nearby the poles in both simulations, but for the orbit I the deposition region is extended from the day to the night side and it is wider compared with the orbit II, limited to the day side. The total mass deposition at the North Hemisphere in the orbit I simulation is larger than in the orbit II, $0.0218$ versus $0.0132$ kg/s. At the South Hemisphere, the asymmetry of the mass deposition is smaller for the orbit I simulation and in both cases it is mainly located on the day side. The mass deposition is larger for the orbit II, $0.03906$ versus $0.03295$ kg/s. The mass deposition on the South Hemisphere is more intense than in the North Hemisphere but the proportion is smaller for the orbit I case, a $40 \%$ of the total mass precipitates at the North Hemisphere, while in the orbit II only a $25 \%$. The total mass deposition in the orbit I simulation is a $5 \%$ larger, consequence of the stronger compression of the BS (the sonic Mach number is higher).

\begin{figure}[h]
\centering
\includegraphics[width=0.6\textwidth]{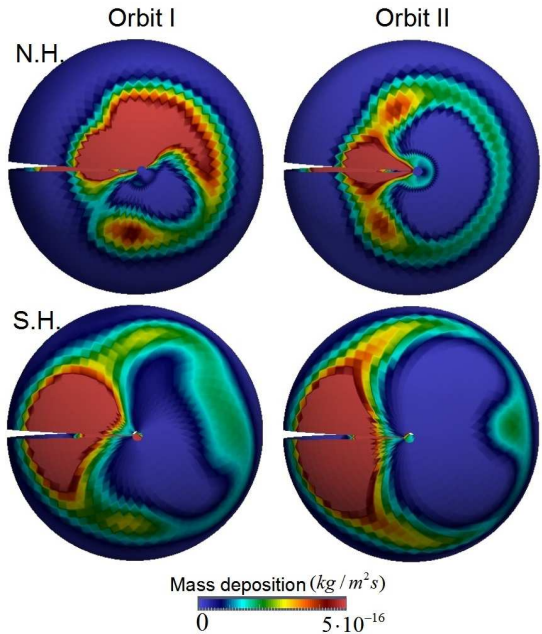}
\caption{Mass deposition on the planet surface for the orbits I and II simulations at the North and South Hemispheres.}
\end{figure}

Figure 6 shows the energy deposition on the Hermean surface, defined as $E = m_{p}v^{2}/2$, and regions with efficient particle sputtering, where $E \geq 2$ eV/p. The largest energy deposition takes place at the South Hemisphere for both simulations, with wide regions of efficient sputtering at middle and high latitudes on the day and night  side. The region with efficient sputtering at the North Hemisphere is limited to the surroundings of the pole on the day side.

\begin{figure}[h]
\centering
\includegraphics[width=1.0\textwidth]{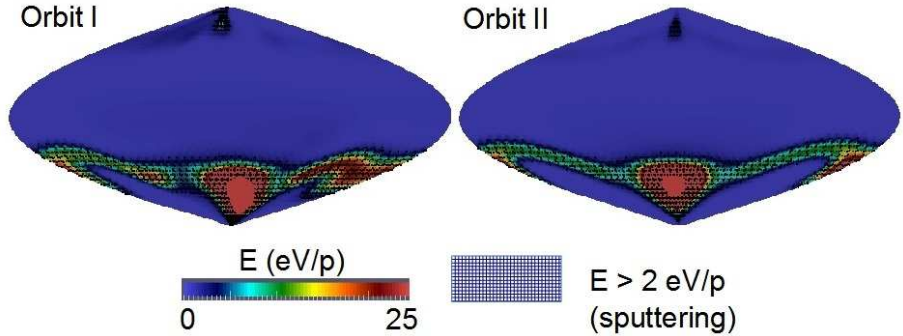}
\caption{Energy deposition on the Hermean surface ($E = m_{p}v^{2}/2$) and regions with efficient particle sputtering ($E \geq 2$ eV/p).}
\end{figure}

\section{Comparison with MESSENGER data}
\label{PS}

In this section we compare the simulations results with MESSENGER magnetometer data and we analyze the density, temperature, pressure, velocity and magnetic field modules obtained in the simulations along the satellite orbit (plotted versus the distance to the satellite closest approach). We include in the graphs the encounter of the satellite with the BS (SI), the magnetopause (MI), the closest approach (CA), and the exit of the satellite from the magnetosphere crossing again the magnetopause (MO) and the BS (SO).

The graphs (7A) to (7D) show that the topology of the magnetic field in the orbit I simulation is comparable to MESSENGER observations, sharing analogue structures in the same positions along the satellite orbit. There are two flattenings of the magnetic field module in MESSENGER data (7A) observed too in the simulations: the first one from $CT = 14:48$ to $14:56$ related with the travel of the satellite along the magnetosheath and the second one from $CT = 14:59$ to $15:05$ that is correlated with the plasma stream in the simulation. The region of main interest is in between MI and CA, second flattening of the magnetic field module, observed when the satellite trajectory crosses the surroundings of the reconnection region. The graphs (7E) to (7I) show that the flattening of the magnetic field module is correlated with a local peak of the density (7E), temperature (7G) and the velocity module (7H) nearby the CA, point out the presence of the plasma stream.

\begin{figure}[h]
\centering
\includegraphics[width=0.8\textwidth]{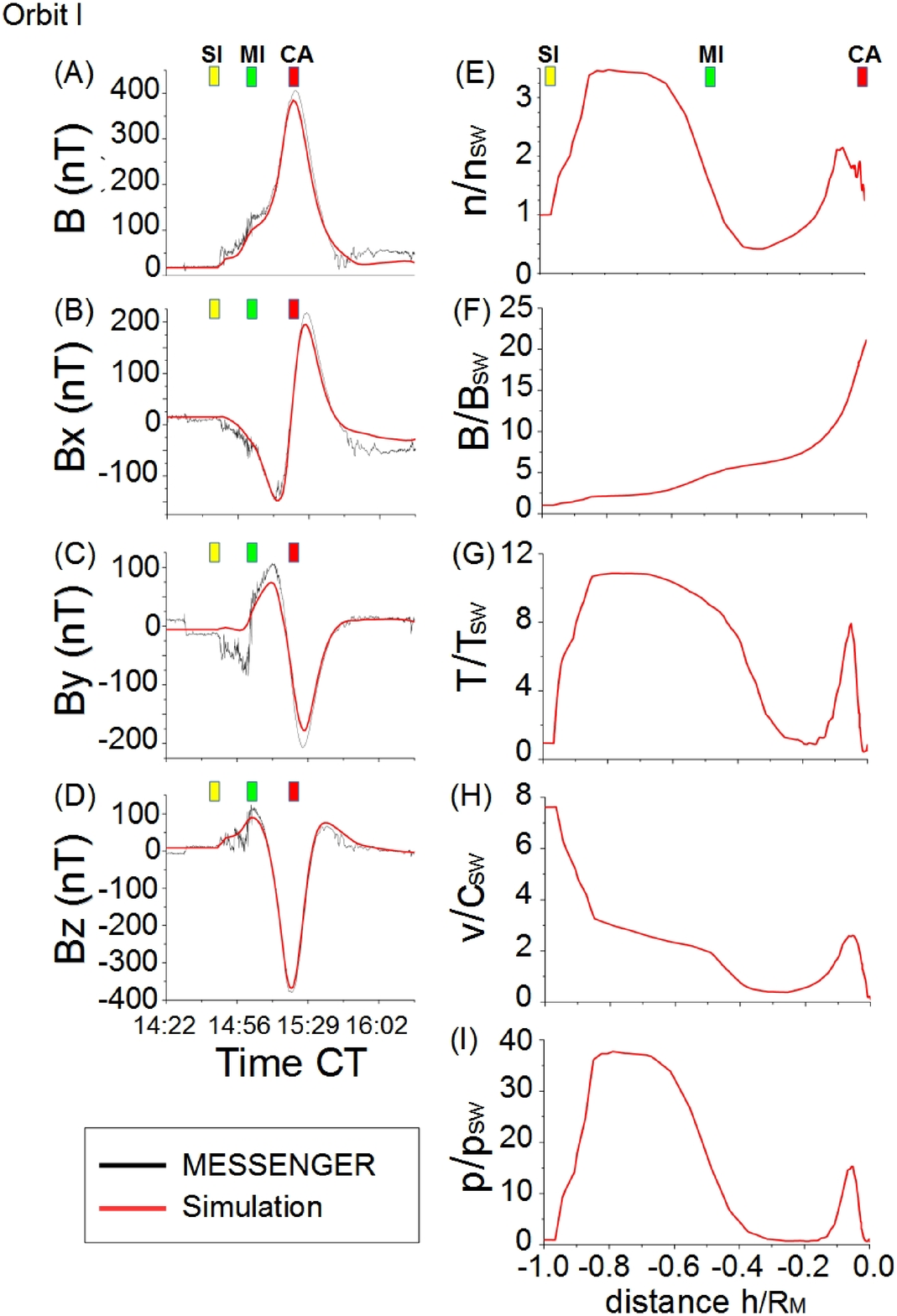}
\caption{(A) to (D) Comparison of the orbit I simulation results (red line) with MESSENGER magnetometer data (black line). (E) Density, (F) magnetic field module, (G) temperature, (H) velocity module and (I) pressure. The density, magnetic field module, temperature and pressure are normalized to the SW values and the velocity to the SW sound speed.}
\end{figure}

We perform the same study for the orbit II simulation. Again the magnetic field topology and the main magnetosphere structures in the simulation and MESSENGER data are similar. There is an averaged drop of the magnetic field module in MESSENGER data (8A) from $CT = 20:31$ to $21:22$ and a profile flattening between $21:31$ and $21:36$ that is correlated with the plasma stream in the simulation. The satellite trajectory in the orbit II crosses a region closer to the reconnection compared with the orbit I, observing a drop of the magnetic field module between MI and CA. There is a flattening of the magnetic field module between MI and CA (8F) correlated with a local maximum of the density (8E), temperature (8G) and velocity module (8H), pointing out the presence of the plasma stream. There is a second peak of the density almost at the satellite CA correlated with a local minimum of the temperature and the velocity, showing the presence of the region with dense and cold plasma accumulated nearby the planet pole before precipitate on the surface.

\begin{figure}[h]
\centering
\includegraphics[width=0.8\textwidth]{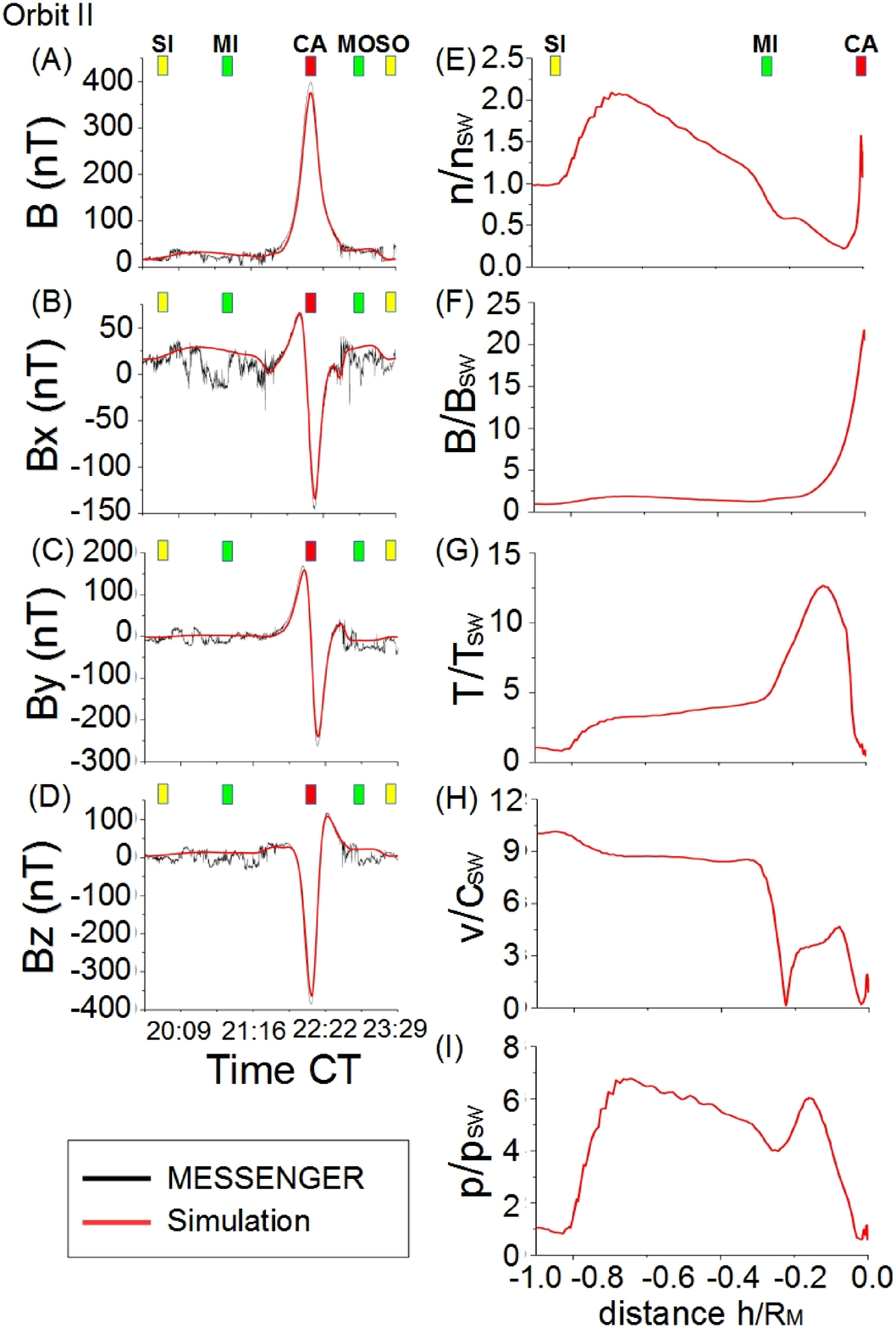}
\caption{Idem Figure 7 for the orbit II simulation.}
\end{figure}

\section{Conclusions}
\label{Conclusions}

The simulations results point out the presence of a plasma stream that links the Hermean magnetosheath with the planet surface at both Hemispheres. The plasma stream originates in the reconnection region between the interplanetary and the Hermean magnetic fields. The plasma in the reconnection region is locally more dense, hotter and it is decelerated before it precipitates along the open magnetic field lines to the planet surface, leading to the formation of a plasma stream. During the precipitation the plasma shows a fast expansion, rarefaction and cooling. In the North Hemisphere, nearby the pole, there is a region of dense and cold plasma where the plasma accumulates before precipitate on the planet surface.

The simulations show that the plasma streams are strongly affected by the location of the reconnection regions and by the asymmetries of the Hermean magnetosphere driven by the interplanetary magnetic field, correlated with a displacement of the regions with the strongest inflows on the planet surface, mass deposition and particle sputtering. The larger compression of the bow shock in the orbit I simulation (the sonic Mach number is a $28 \%$ higher) leads to an enhancement of the particle flux on the planet surface, almost a $5 \%$ larger. The solar wind dynamic pressure in the orbit I simulation is $3.125 \cdot 10^{-9}$ Pa versus $2.16 \cdot 10^{-9}$ Pa in the orbit II simulation.

The topology of the magnetic field in the simulation is comparable to MESSENGER observations, sharing analogue structures in the same positions along the satellite orbit. The discrepancies observed in the graphs 7C and 8B are related with the model limitations; in the graph 7C the West rotation of the $B_{y}$ component is not well reproduced due to the lack of resolution leading to a smooth out of the gradient, while in the graph 8B the $B_{x}$ shows non negligible oscillations during the transit of the satellite from the BS to the MI so there is a mismatch of this component tendency because we perform a steady state simulation. The simulations show that the satellite crosses the plasma stream in a region where the profile of the magnetic field module is flattened due to the proximity of the reconnection region, correlated with a local maximum of the density, temperature and velocity. The orbit II simulation shows that the satellite crosses a region of plasma accumulation nearby the North pole before it precipitates on the planet. 
 
The present simulations results are compatible and complementary with previous observational studies of particle fluxes on the Hermean surface and and magnetosheath plasma depletion \cite{2013JGRA..118.7181G,2013AGUFMSM24A..03D,2007SSRv..132..433K}. The plasma depletion layer is not resolved in the present simulations as an independent structure of the magnetosheath due to the lack of resolution of the model, but the simulation features are compatible with the observations in the transition between the magnetosheath and the magnetopause. The magnetic pile-up in the planet dayside is not observed in the simulations because the numerical resistivity is several orders larger than the real plasma conditions so the reconnection between the interplanetary and the Hermean magnetic field is instantaneous, but the model can predict the location and foreseen the essential role of the reconnection region in the magnetosheath plasma depletion. The results are also compatible with the observations of protons precipitation from the magnetosheath along the Hermean cusp, accelerated and transported to the Hermean surface \cite{2014JGRA..119.6587R}.The global magnetosphere structures are similar to numerical simulation performed by other author using different numerical schemes \cite{2008Icar..195....1K,2007AGUFMSM53C1412T}. We focus the study in the properties of the plasma stream from its origin in the magnetosphere reconnection region to the planet surface, as well as the consequences of an enhanced bow shock compression in the mass deposition and particle sputtering, continuation of a previous communication devoted to analyze the key role of the interplanetary magnetic field orientation on the fluxes toward the Hermean surface \cite{2015PandSS..119..264V}.

\section{Aknowledgments}
The research leading to these results has received funding from the European Commission's Seventh Framework Programme (FP7/2007-2013) under the grant agreement SHOCK (project number 284515). The MESSENGER magnetometer data set was obtained from the NASA Planetary Data System (PDS) and the values of the solar wind hydrodynamic parameters from the NASA Integrated Space Weather Analysis System.

\section*{References}

\bibliography{mybibfile}

\end{document}